\documentclass[
]{revtex4-2}
\usepackage{amssymb}
\usepackage{amsmath}

\usepackage{makecell}
\usepackage{cellspace} 
\usepackage{chemformula}
\usepackage[symbol]{footmisc}
\setcitestyle{comma}
\setcitestyle{square}
\usepackage{comment}

\usepackage{caption} 
\usepackage{bm}
\usepackage{hyperref}
\usepackage{xcolor,soul}

\usepackage{chemformula}

\begin{document}

\title{Learning continuous state of charge dependent thermal decomposition kinetics for Li-ion cathodes using Kolmogorov-Arnold Chemical Reaction Neural Networks (KA-CRNNs)}

%\author{{\large Benjamin C. Koenig, Sili Deng\footnote[1]{Corresponding Author. \\  \emph{E-mail Address:} silideng@mit.edu (S. Deng)}}\\[10pt]
%        {\footnotesize \em Department of Mechanical Engineering, Massachusetts Institute of Technology, Cambridge, MA 02139, USA}\\[-3pt]}
    
\author{Benjamin C. Koenig}
\affiliation{Department of Mechanical Engineering, Massachusetts Institute of Technology, 77 Massachusetts Ave, Cambridge, MA 02139, United States.}

\author{Sili Deng}
\email{Corresponding author, silideng@mit.edu}
\affiliation{Department of Mechanical Engineering, Massachusetts Institute of Technology, 77 Massachusetts Ave, Cambridge, MA 02139, United States.}
\date{\today}% It is always \today, today,

\begin{abstract}

Thermal runaway in lithium-ion batteries is strongly influenced by the state of charge (SOC). Existing predictive models typically infer scalar kinetic parameters at a full SOC or a few discrete SOC levels, preventing them from capturing the continuous SOC dependence that governs exothermic behavior during abuse conditions. To address this, we apply the Kolmogorov-Arnold Chemical Reaction Neural Network (KA-CRNN) framework to learn continuous and realistic SOC-dependent exothermic cathode-electrolyte interactions. We apply a physics-encoded KA-CRNN to learn SOC-dependent kinetic parameters for cathode–electrolyte decomposition directly from differential scanning calorimetry (DSC) data. A mechanistically informed reaction pathway is embedded into the network architecture, enabling the activation energies, pre-exponential factors, enthalpies, and related parameters to be represented as continuous and fully interpretable functions of the SOC. The framework is demonstrated for NCA, NM, and NMA cathodes, yielding models that reproduce DSC heat-release features across all SOCs and provide interpretable insight into SOC-dependent oxygen-release and phase-transformation mechanisms. This approach establishes a foundation for extending kinetic parameter dependencies to additional environmental and electrochemical variables, supporting more accurate and interpretable thermal runaway prediction and monitoring. 

\end{abstract}

\maketitle

\section{Introduction} \label{intro}
Lithium-ion batteries are a key technology for meeting evolving global energy demands. As their prevalence continues to rise, it has become increasingly important to properly quantify and control their safety against catastrophic thermal runaway. Thermal, electrical, and mechanical abuse all provide pathways that lead to accelerating exothermic behavior, heat-releasing internal short circuits, and even extreme combustion events \cite{wang_review_2019}. Reliable and robust models for these phenomena are a key piece in thermal runaway prevention and mitigation.

Differential scanning calorimetry (DSC) is widely used to develop kinetic models that describe component-specific reaction pathways, from standard operating conditions to extreme thermal events. These can include cathode decomposition \cite{wang_thermal_2021, koenig_accommodating_2023, koenig_uncertain_2024}, solid electrolyte interphase degradation and regeneration \cite{kriston_quantification_2019, ren_model-based_2018}, and anode-electrolyte interactions \cite{ren_model-based_2018, koenig_comprehensive_2025}, among other minor components and major cross-component interactions. The Kissinger method is a traditional technique for the extraction of kinetic model parameters from these component-scale DSC datasets \cite{kissinger_variation_1956, macneil_reactions_2002, kriston_quantification_2019}, although it has been shown generally \cite{vyazovkin_ictac_2011, vyazovkin_kissinger_2020} as well as quantitatively in a specific lithium-ion battery kinetic modelling study \cite{koenig_accommodating_2023} that the simplifying assumptions required for its use severely limit its accuracy and capability when attempting to model realistic chemical pathways. Chemical Reaction Neural Networks (CRNN), originally proposed for general autonomous kinetic model discovery \cite{ji_autonomous_2021}, have been leveraged extensively as a high-accuracy alternative with the capability for complex and realistic chemical physics inference for battery thermal runaway models via their combination of machine learning optimization with the Arrhenius and mass action laws \cite{koenig_accommodating_2023, koenig_uncertain_2024, koenig_comprehensive_2025, bhatnagar_chemical_2025, zhang_chemical_2025}. {In fact, a detailed head-to-head comparison of the CRNN approach and the Kissinger approach in a thermal runaway-specific context \mbox{\cite{koenig_accommodating_2023}} revealed significant improvements both in model accuracy to the experimental data and in model accuracy to the underlying physics when switching to the CRNN approach.} However, existing CRNN-based models treat kinetic parameters as fixed scalars and therefore cannot represent dependencies on external conditions that are not already represented in the Arrhenius and mass action laws.

State of charge (SOC) is a particularly important external factor because it strongly influences both global and component-scale thermal runaway severity. Increased SOC elevates internal short circuit energy release \cite{feng_thermal_2015}, and similarly cathode materials exhibit greater exothermicity at higher SOCs (higher delithiation levels), both in electrolyte-containing \cite{kvasha_comparative_2018, stenzel_thermal_2019} and electrolyte-free settings \cite{kvasha_comparative_2018, cui_navigating_2025}. Despite this well-established dependence, most kinetic modeling efforts focus exclusively on the 100\% SOC condition to generate conservative worst-case safety predictions \cite{wang_thermal_2021, koenig_comprehensive_2025, ren_model-based_2018}. Although practical, this approach cannot capture the continuous variation in exothermic behavior that occurs across realistic operating and storage conditions, limiting the predictive capability of such models at realistic, intermediate SOC conditions.

Recent attempts to introduce SOC sensitivity into kinetic modeling have followed two main directions. One approach applies Kissinger-type parameter fitting independently at a limited number of discrete SOC values \cite{karmakar_state--charge_2024, he2023investigation}. These studies provide valuable insight into overall SOC trends but retain the limitations of the underlying analytical framework and do not produce a continuous representation suitable for real-time prediction. A second, more recent approach employs CRNNs with accelerating rate calorimetry (ARC) data to infer SOC-sensitive global heat release and gas evolution \cite{zhang_chemical_2025}. This method achieves strong predictive performance on global thermal runaway trends and is able to accurately model the SOC behavior of the full cell. However, the low resolution and mixed-component nature of ARC data restricts the identifiability of component-specific pathways and intermediate species, and requires the use of lumped reactants with limited correlation to realistic decomposition phenomena. As a result, the inferred models cannot provide mechanistic interpretability or component-resolved decomposition kinetics.

Parallel DSC experimental studies sweeping the SOC at fine resolution have revealed a striking phenomenon: the total heat released by mixed cathode-electrolyte samples increases gradually with SOC until a critical value is reached, beyond which the heat release and DSC peak shapes change abruptly and thermal stability rapidly drops \cite{cui_navigating_2025}. This critical SOC appears to identify a regime transition with significant implications for battery operation, storage, and safety management optimization. Prior work has associated this behavior with sharply increasing lattice oxygen release from nickel-rich cathodes at similar elevated SOCs \cite{xiong_measuring_2017, jung_understanding_2014, jung_chemical_2017, bak_correlating_2013, jung_oxygen_2017}. This oxygen release accelerates exothermic decomposition and electrolyte oxidation, suggesting a mechanistic link between SOC, oxygen evolution, and thermal runaway propensity. However, no existing kinetic framework provides a continuous and fully interpretable SOC-dependent representation of these processes. 

The recently proposed Kolmogorov-Arnold Chemical Reaction Neural Network (KA-CRNN) framework \cite{koenig2025kolmogorov} offers a means to address this gap. KA-CRNN integrates the mechanistic structure of CRNNs with the functional expressiveness of Kolmogorov-Arnold Networks \cite{liu_kan_2024, koenig_leankan_2025} and Neural ODE formulations \cite{koenig_kan-odes_2024, koenig_chemkans_2025, chen_neural_2019}. The resulting architecture enables Arrhenius parameters to vary smoothly with SOC while remaining anchored in physically {reasonable} reaction networks. In this work, we construct a KA-CRNN model based on literature-supported cathode decomposition and electrolyte oxidation pathways and apply it to nickel cobalt aluminum, nickel manganese aluminum, and nickel manganese cathodes. By learning kinetic parameters directly from DSC measurements across a range of SOC values, the KA-CRNN produces a continuous, mechanistically grounded model that captures the experimentally observed critical SOC behavior \cite{cui_navigating_2025} and provides component-specific predictive capability for exothermic reactions relevant to thermal runaway under realistic operating and storage conditions.

\section{Methodology} \label{sec:methods}

Section \ref{sec:method-review} of the methodology reviews the literature and selects a reasonable model form to capture key SOC dependence behavior in cathode-electrolyte decomposition models. Then, Sec. \ref{sec:methods_KACRNN} introduces the KA-CRNN formulation to learn the parameterized behavior of this model. We finally introduce the training methodology and datasets in Sec. \ref{sec:methods_data}.

\subsection{Cathode-electrolyte decomposition modeling overview}\label{sec:method-review}

Typical cathode decomposition models developed in the thermal runaway literature, both via analytical techniques \cite{wang_thermal_2021, ren_model-based_2018, feng_thermal_2015} and CRNNs \cite{koenig_accommodating_2023, koenig_uncertain_2024, koenig_comprehensive_2025}, report one to three decomposition steps that correspond to global phase change reactions. In NCM cathodes for example, an established pathway involves the pristine layered structure, two sequential spinel structures, and the final rock salt \cite{bak_structural_2014, wang_thermal_2021, koenig_accommodating_2023}. Such modeling efforts deliver compact models with scalar parameters by limiting inference to a single SOC only. A well-known product of these decomposition reactions is oxygen gas \cite{bak_correlating_2013, sharifi2019oxygen, xiong_measuring_2017}, which can oxidize the electrolyte and lead to severe exothermic behavior \cite{belharouak2006thermal, ping2014thermal} including fire and explosion \cite{wang2012thermal, henriksen2019explosion}. 

The key recent result motivating the modeling effort of this work was the observation of a critical SOC in cathode-electrolyte DSC samples above which there is a sharp increase in exothermic behavior accompanied by a sudden narrowing of the heat release peak \cite{cui_navigating_2025}. This observation was made identically across 15 different nickel-rich cathode chemistries, indicating its prevalence in modern battery systems. The seemingly discontinuous nature of this abrupt shift in thermal decomposition behavior, with otherwise well-behaved and smooth trends at lower SOCs, suggests an underlying physical process that similarly responds suddenly to high SOCs rather than gradually at all values. Proper treatment of this underlying process could be the key to enabling highly informative, physics-based kinetic models that respond continuously to the SOC and enable higher-performance monitoring systems. We discuss the chemical and electrochemical origins of this behavior in the two subsections below, starting with the SOC dependence of oxygen evolution from the cathode active material during phase transition and then continuing to electrolyte oxidation and combustion pathways that are highly coupled to the released oxygen.

\subsubsection{Cathode oxygen evolution dependence on lithiation}\label{sec:method-cath-theory}

We begin with discussion of cathode electrochemical decomposition behavior during standard cycling operation, where phase transitions similar to those experienced during thermal abuse are commonly observed. Prior study on the irreversible cycling transformation of surface lattice structures for NCM cathodes \cite{jung_understanding_2014} concluded that the type of phase change decomposition experienced at the surface of the cathode depends on the voltage condition: at elevated but still relatively low voltages corresponding to a half or more delithiated cathode, the observed cycling phase change is to the spinel structure, with little to no oxygen release. Heavily delithiated voltages near 4.8V, meanwhile, favor complete decomposition to the rock salt phase due to the highly oxidative environment. This high-voltage decomposition notably releases significant amounts of oxygen. Modeling of this surface reconstruction \cite{zhuang_theory_2022} has similarly concluded that the key degradation behavior (and implicitly the oxygen release characteristics) are governed by an electrochemical voltage cutoff rather than a smooth trend: below 4.4V the model computed zero surface reconstruction or capacity fade, while above 4.4V standard cycling led to phase change and degradation.

On the thermal side of the literature, TGA-MS studies \cite{xiong_measuring_2017} on an array of uncoated, coated, single crystal, and polycrystalline NCM cathodes of varying nickel ratios charged to 4.2, 4.4, and 4.6V arrived at a similar conclusion: little to no oxygen evolution was measured at 4.2V, with significant increases at 4.4V and 4.6V. Another gas evolution study \cite{bak_correlating_2013} corroborates these results with its own experimental data and further proposes a set of two-step reactions for nickel-rich NCM cathodes at varying lithiation levels. We reproduce the lowest-charge (lithiated pristine cathode) pathway first,

\begin{align}
    & \text{Li}_{1/2}(\text{M}^{3.5+})\text{O}_2 \ch{->} \frac{1}{2}(\text{Li}(\text{M}^{3.5+})_{2}\text{O}_4), \\
    & \frac{1}{2}(\text{Li}(\text{M}^{3.5+})_{2}\text{O}_4) \ch{->} \frac{1}{2}(\text{Li}(\text{M}^{3.5+})_{2}\text{O}_3) +\frac{1}{4}O_2,
\end{align}
with the top reaction representing the phase change from the layered cathode to the spinel phase, and the bottom reaction representing the spinel phase to rock salt decomposition. Notable here is the lack of oxygen production during the first decomposition step, and the minor oxygen production at the second step. In contrast, the highest-charge pathway (delithiated pristine cathode), reproduced again from \cite{bak_correlating_2013}, is as follows:

\begin{align}
    & \text{Li}_{1/10}(\text{M}^{3.9+})\text{O}_2 \ch{->} \frac{11}{30}(\text{Li}_{3/11}(\text{M}^{2.84+})_{30/11}\text{O}_4) + \frac{4}{15}O_2, \\
    & \frac{11}{30}(\text{Li}_{3/11}(\text{M}^{2.84+})_{30/11}\text{O}_4) \ch{->} \frac{11}{30}(\text{Li}_{3/11}(\text{M}^{2.84+})_{30/11}\text{O}_3) +\frac{11}{60}O_2,
\end{align}
where similar overall steps but adjusted stoichiometry and charge balances lead to significant oxygen release in the first step and an $80\%$ increase in total production. We gather from this series of electrochemical and thermal cathode studies that there is significant experimental justification for a sharp increase in oxygen released from the cathode at high SOCs  (or correspondingly, high voltages). In the next section, we connect this oxygen evolution to exothermic behavior at the electrolyte, and propose a KA-CRNN model form to bring these behaviors together in a cohesive numerical model that can explain the critical SOC phenomenon using these real chemical physics behaviors.

\subsubsection{Electrolyte oxidation and proposed model form} \label{sec:method-elec-model}

Further theory is needed to correlate this sudden oxygen release increase with the critical exothermic safety behavior observed in \cite{cui_navigating_2025}, where DSC samples included cathode samples mixed with liquid electrolyte. Electrolyte oxidation has been widely studied in chemical and electrochemical settings \cite{jung_oxygen_2017, jung_chemical_2017, zhang_revealing_2020}, where the electrochemical pathway via carbonate dehydrogenation occurs relatively slowly at extreme voltages, and the chemical oxidation pathway that relies on cathode-released oxygen is known to increase in rate suddenly at still elevated but relatively lower voltages \cite{jung_oxygen_2017}. This connection between cathode-evolved oxygen and electrolyte oxidation or combustion, and its significant impact on practical-scale thermal runaway, has been well-studied \cite{belharouak2006thermal, ping2014thermal, wang2012thermal, henriksen2019explosion, xiang_thermal_2009, wang_first-principles_2007}, and provides a strong link between the sudden increase in oxygen evolution described in Sec. \ref{sec:method-cath-theory} and the seemingly related sudden increase in cathode-electrolyte heat release \cite{cui_navigating_2025}. We theorize in this work that proper, continuous treatment of the evolved oxygen stoichiometry is the key to developing a truly global, SOC-varying model for this cathode-electrolyte exothermic behavior, and may help to lay the groundwork for extended applications in other battery materials.

We briefly remark that the results reviewed here are not fully consistent on whether the oxygen evolution in cathode materials depends on the SOC or voltage. {This is especially relevant given the literature on algorithmic approaches for SOC estimation. While not in the scope of the current work as we leverage existing datasets, such algorithms apply various approaches such as Kalman filters, neural networks, and coulomb counting to estimate the SOC \mbox{\cite{hou2023robust}}.} Returning to the question of SOC vs. voltage as the primary signal, we remark that some work has even suggested that the true answer is neither, and that the differential capacity (dq/dV) is the fundamental predictor of such behavior \cite{jung_oxygen_2017}. While perfect correlations do not exist due to transient effects and overpotentials, rough conversions of all metrics via charging curves \cite{cui_navigating_2025} reveal that all effects discussed here occur in similar ranges across all three variables. For the purposes of this study, we select the SOC as the dependent variable, but note that with controlled experiments the voltage and differential capacity are also suitable choices and that further investigation here may reveal new insights. 

General conclusions of the above review are as follows:
\begin{enumerate}
    \item Thermal decomposition of layered nickel-rich cathodes occurs in at least two steps, through the spinel and rock salt phases. Proper modeling efforts must account for variation in oxygen release across these steps.
    \item The bulk of the oxygen release occurs in the second step, in quantities generally agreed upon in the literature to vary significantly with the SOC.
    \item The released oxygen reacts strongly with the electrolyte, producing significant exothermic behavior. We find no consensus in the literature suggesting that the electrolyte oxidation kinetics vary significantly with the SOC.
    \item Reported voltage and SOC thresholds in the literature indicate that the critical SOC for thermal safety \cite{cui_navigating_2025} occurs at the same SOC as sudden increases in cathode oxygen evolution. This suggests that a model properly accounting for the SOC-based variation in cathode oxygen evolution will inherently be capable of predicting SOC-dependent exothermic effects.
\end{enumerate}
Following from these points, we propose the following simple model form for general nickel-rich exothermic cathode decomposition:
\begin{align}
    C_{l} & \ch{->[R1]} C_{s} +\Delta H_1, \label{eq:r1}\\ 
    C_{s} & \ch{->[R2]} C_{r} + \nu O_2 +\Delta H_2, \label{eq:r2}\\
    O_2 + C_{e} & \ch{->[R3']} C_{prod} + \Delta H_3'.\label{eq:r3}
\end{align}
Here $C_l$ is the pristine layered cathode structure, $C_s$ is the spinel structure, and $C_r$ is the rock salt for a generic nickel-rich material. $O_2$ is also present in the system, as is $C_{prod}$, the product of the reaction between $O_2$ and the electrolyte $C_e$. Enabling even a simple kinetic model like this to respond to the SOC presents a challenge. The rate parameters of the electrochemically-dependent R1 and R2 differ for each studied cathode chemistry, and also must respond continuously to the SOC. R3', meanwhile, represents the chemical electrolyte oxidation pathway that occurs regardless of cathode or SOC, as indicated by the prime notation. Similarly, the heat released from the first two reactions is SOC dependent, while the prime $\Delta H_3'$ is a fixed constant based on the electrolyte chemistry. 

\begin{figure}
\centering
\includegraphics[width=\linewidth]{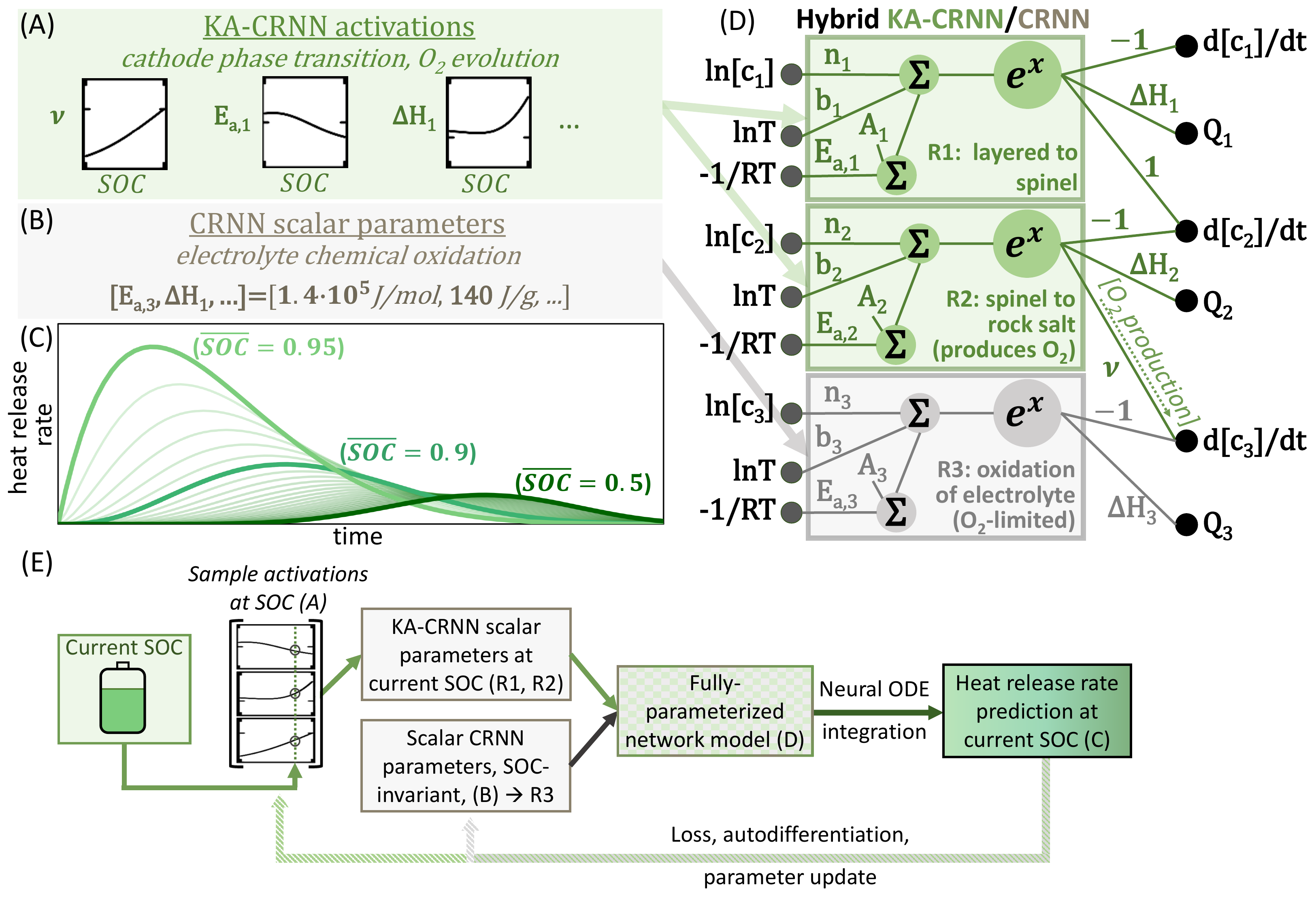}
	\caption{Overview of KA-CRNN framework for SOC-dependent thermal runaway kinetics. (A) KA-CRNN kinetic parameters, represented as continuous functions of SOC. (B) Standard CRNN kinetic parameters, represented as fixed scalar values without SOC dependence. (C) Example KA-CRNN prediction illustrating the continuous variation of heat-release behavior with SOC.  (D) Complete KA-CRNN architecture used in this study. Reactions R1 and R2 are defined by KA-CRNN parameters as in panel (A) and are identified separately for each cathode material. Reaction R3 uses scalar CRNN parameters as in panel (B) and is learned once for all cathodes, but its effective behavior remains SOC dependent through its reliance on O$_2$ production from R2. (E) Overall flowchart depicting the forward solve and backwards parameter update process. Note references within (E) to subfigures (A-D) for clarity.}
	\label{fig:fig1}
\end{figure}

Returning to the four constraints listed above, we note that the the two-step cathode decomposition in R1 and R2 satisfies conclusion (1), the SOC dependent coefficient $\nu$ in R2 satisfies (2), the presence of R3 satisfies (3), and KA-CRNN training of the oxygen evolution reaction using critical SOC data will implicitly enforce (4). The continously varying kinetic parameters of R1 and R2, especially the oxygen stoichiometric coefficient $\nu$, are the key modeling innovation that enable this concise yet still strongly physics-encoded reaction model. {Specifically, the learnable SOC variation of $\nu$ enables it to adjust, assumption-free, from minimal oxygen evolution at low SOCs to large oxygen evolution at high SOCs, providing a direct connection to the oxygen release experimental conclusions presented in Sec. {\ref{sec:method-cath-theory}}.} To accurately learn these continuous kinetic parameters, we leverage KA-CRNNs as discussed in the next subsection and shown in Fig. \ref{fig:fig1}.

\subsection{Physics-encoded KA-CRNN with O$_2$ release and electrolyte oxidation}\label{sec:methods_KACRNN}
The KA-CRNN approach \cite{koenig2025kolmogorov} first involves the definition of a complete CRNN via the Arrhenius and mass action laws \cite{ji_autonomous_2021}, as is prevalent in the literature for thermal runaway kinetics \cite{koenig_accommodating_2023, koenig_uncertain_2024, bhatnagar_chemical_2025, koenig_comprehensive_2025}. Then, the subset of kinetic parameters defined in the CRNN for R1 and R2 are re-parameterized using KAN activations to incorporate their continuous functional dependnce on an external factor, here the SOC. The CRNN reaction system is as defined in Eqs. \ref{eq:r1} through \ref{eq:r3}. The reaction rates can be computed as
\begin{equation}\label{eqn_rate_log}
r_i=\text{exp}(n_{i} \text{ln}[c_i] + \text{ln} A_i + b_i \text{ln} T - E_{a,i}/RT)
\end{equation}
for the $i^{th}$ reaction, where $i=1:3$. $c_1$ is the normalized mass of the layered structure corresponding to $C_l$, while $c_2$ is the spinel phase ($C_s$) and $c_3$ is the oxygen (O$_2$). {The normalized mass of the electrolyte corresponding to $C_e$ is fixed at unity due to the oxygen-limited reaction, thus it has a constant zero effect on Eq. {\ref{eqn_rate_log}}. $C_r$ and $C_{prod}$ are exclusively reaction products that do not appear as reactants in any reactions studied here. Thus, while their formation remains implicitly modeled by the reaction rates studied here, we do not explicitly track their normalized masses to reduce extraneous computation. This omission does not affect any of the learned reaction rates or heat release profiles, as these species have no impact on these quantities as per the formulation in Eq. {\ref{eqn_rate_log}}.} The individual species production and consumption rates are

\begin{align}
\frac{dc_1}{dt}&=-r_1, \ \ \ \text{ [layered phase]}\label{eqn_species_rate_1} \\
\frac{dc_2}{dt}&=-r_2+r_1, \ \ \ \text{[spinel phase]]}\label{eqn_species_rate_2} \\
\frac{dc_3}{dt}&=-r_3+\nu r_2. \ \ \ \text{ [evolved $O_2$]}\label{eqn_species_rate_3}
\end{align}
 Finally, the key exothermic heat release term is defined using the individual reaction enthalpies $\Delta H_i$ as

\begin{equation}\label{eqn_heat}
Q_{tot}=\sum_i Q_{i} = \sum_i \Delta H_i r_i.
\end{equation}

The rates reported in Eqs. \ref{eqn_species_rate_1} through \ref{eqn_heat} can be seen in the one-to-one mapping on the output side of Fig. \ref{fig:fig1}(D). {We briefly clarify here that these are not the elementary reactions describing molecular-scale behavior, but are rather a series of global phase change and oxidation reactions that, based on previous work in the literature, are sufficient to capture the global trends of interest to the thermal runaway modeling community.} {In a traditional CRNN approach} \cite{ji_autonomous_2021, koenig_accommodating_2023, koenig_comprehensive_2025}, {the series of complete kinetic parameters used in Eq. {\ref{eqn_rate_log}} and Eq. {\ref{eqn_heat}} (which we remark is fully defined by the one-to-one mapping of parameters in Fig. {\ref{fig:fig1}}(D)) would be integrated forward using the Neural Ordinary Differential Equations approach \mbox{\cite{chen_neural_2019}}, and then fine-tuned via MSE loss to iteratively learn the set of real scalar model parameters that best fits the data. Here, however, we require a model that responds continuously to the state of charge, thus the scalar-valued CRNN approach is not sufficient.}

Kinetic parameter responses to the SOC are encoded via univariate KAN activations \cite{liu_kan_2024, koenig_kan-odes_2024}, as originally demonstrated in a proof-of-concept KA-CRNN \cite{koenig2025kolmogorov}. The i$^{th}$ SOC-varying kinetic parameter is defined as

\begin{equation} \label{kan_param}
    p_i\left(\text{SOC}\right) = \text{KAN}\left(\text{SOC}, \bm{\theta}_i\right) =
    {\phi}_i\left(\text{SOC}\right).
\end{equation}
{Here, $p_i$ is the learned continuous representation of the $i^{th}$ SOC-varying kinetic parameter, a full list of which is available in Table {\ref{tab:params}}. Each $i^{th}$ parameter is a univariate continuous function of the SOC, as illustrated in Fig. {\ref{fig:fig1}}(A). This encoding retains the overall kinetic model structure of Fig. {\ref{fig:fig1}}(D) as would be applied with a standard CRNN, but rather than scalar parameters as in a standard CRNN the parameters now change with the SOC.}

{In more detail, the middle term in Eq. {\ref{kan_param}} is defined such that each distinct $i^{th}$ parameter $p_i$ is encoded using the same overall KAN structure (note the function KAN does not include a subscript $i$), although with a uniquely learned set of network parameters $\bm{\theta}_i$. On the right hand side, the KAN combined with the $i^{th}$ parameters makes up a single fully interpretable activation function ${\phi}_i$. This simple encoding facilitates convergence and human-interpretable meaning by isolating the specific effect of the SOC on each individual parameter and retaining extremely sparse representations using single activation functions only. These ${\phi}_i$ are defined using the Chebyshev KAN formulation \mbox{\cite{ss2024chebyshev}} as}

\begin{align}
    &\phi_{i} \left(SOC \right) = \sum_{n=0}^{N} w^{\psi}_{i, n} \cdot \psi_n\left(\text{SOC}\right),\label{eq:ChebyKAN} \\ 
    &\psi_n(\text{SOC}) = \text{cos}(n \cdot \text{arccos}(\text{SOC})).  \label{eq:ChebyPoly}
\end{align}
{In Eq. {\ref{eq:ChebyKAN}}, we define an individual activation ${\phi}_i$ corresponding to a single SOC-varying kinetic parameter as a sum of Chebyshev polynomials $\psi_n$ up to degree $N$. The scaling weights $w^{\psi}_{i,n}$ of each polynomial constitute the learnable parameters of the KAN, that is $\bm{\theta}_i$ = $[w^{\psi}_{i,1},...w^{\psi}_{i,n}]$. Meanwhile, the functional form of each Chebyshev polynomial $\psi_n$ is provided in Eq. {\ref{eq:ChebyPoly}}. Thus with the fully converged model (i.e., all $w^{\psi}_{i,n}$ are converged after training), we can produce the CRNN structure of Fig. {\ref{fig:fig1}}(D) at any arbitrary SOC simply by solving Eq. {\ref{eq:ChebyKAN}} at that SOC. This pipeline is shown on the left hand side of Fig. {\ref{fig:fig1}}(E), where the current SOC is fed into the KA-CRNN activations (identically, fed into Eq. {\ref{eq:ChebyKAN}}) to produce the set of scalar parameters used in the CRNN structure of Fig. {\ref{fig:fig1}}(D). } We use the continuous Chebyshev polynomial formulation here rather than the standard gridded basis KAN formulation to accommodate the relatively sparse training data.

{Each studied cathode material has all kinetic parameters for its phase transition reactions (R1 and R2) defined using this formulation, with cathode-specific collections of parameters $\bm{\theta}$ that are simply the set of that cathode's parameter-specific arrays $\bm{\theta}_i$. In addition, a standard set of scalar CRNN parameters is used for the universal electrolyte oxidation model (R3). As shown in Figs. {\ref{fig:fig1}}(B) and {\ref{fig:fig1}}(E), the R3 parameters do not vary with the SOC. A summary of this split is available in Table {\ref{tab:params}}. Thus, for a given cathode at a given SOC, we first convert the KAN-based parameters $\bm{\theta}$ into scalar values using Eq. {\ref{eq:ChebyKAN}}, and then construct the resulting standard CRNN of Fig. {\ref{fig:fig1}} for forward solving by combining these values with the scalar R3 values, as in the traditional CRNN approach} \cite{ji_autonomous_2021, koenig_accommodating_2023}. {This fusion of the KAN approach and the CRNN approach enables a fully continuous, assumption-free representation of the kinetic parameters' responses to the SOC, while retaining the strongly physics-encoded CRNN backbone through which the parameters are integrated to reconstruct the training data. For all parameter activations in this work, we use $N=10$, which when counting the zeroth degree polynomial term, results in eleven KAN parameters used to learn each true kinetic parameter (121 KAN parameters used to fully learn the 11 real kinetic parameters in each cathode's 2-step decomposition pathway). $N=10$ was found to be a high-performing choice in this framework thanks to its balance between model sparsity, preventing overfitting and spurious oscillatory behavior in the learned parameters between sample data points; and model detail, enabling effective inference of the relatively refined critical SOC behavior. To provide a starting point for future study, we remark that the experimentally selected value of $N=10$ is similar to the nine SOC samples per model (shown in Fig. {\ref{fig:data}}). A complete visualization of the model framework is available in Fig. {\ref{fig:fig1}}.}

\subsection{KA-CRNN training: NMA, NM, and NCA cathodes} \label{sec:methods_data}
\begin{figure}
\centering
\includegraphics[width=\linewidth]{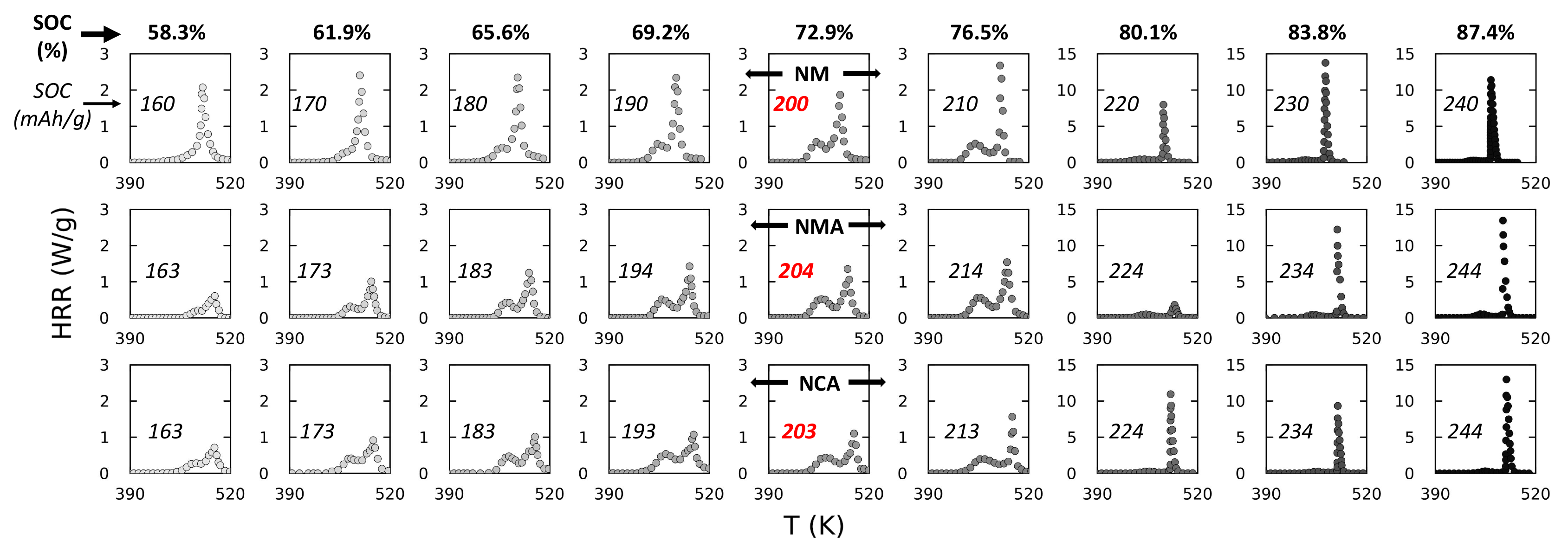}
	\caption{DSC training data reproduced from \cite{cui_navigating_2025}. Rows are NM, NMA, and NCA cathode materials from top to bottom. The y-axis scale changes beginning at the seventh column to improve visibility of the substantially increased heat release at elevated SOCs. {SOCs are reported globally in percentages for each column, as well as locally in mAh/g for each subplot to accommodate the different maximum capacities of each chemistry.}}
	\label{fig:data}
\end{figure}

To demonstrate the versatility and robustness of the KA-CRNN technique, we train on three separate nickel-rich cathode datasets reproduced from \cite{cui_navigating_2025}, all of which experienced the critical SOC phenomenon: nickel cobalt aluminum (NCA), nickel manganese (NM), and nickel manganese aluminum (NMA). {These datasets were collected {\cite{cui_navigating_2025}} from samples containing rinsed and dried cathode powders injected with LP57 electrolyte, at a heating rate of 2$^{\circ}$C/min on a Netzsch STA 449 system.} Each dataset contains nine DSC experiments at varying SOCs. {These SOC values were calibrated in the original experimental study \mbox{\cite{cui_navigating_2025}} to be the same percentage of lithium removal for all three cathodes: 87.4$\%$, 83.8$\%$, 80.1$\%$, 76.5$\%$, 72.9$\%$, 69.2$\%$, 65.6$\%$, 61.9$\%$, and 58.3$\%$. These roughly correspond to specific capacities of 240, 230, 220, 210, 200, 190, 180, 170 and 160mAh/g, although the specific charge values differ by a few percentage points across the three cathode chemistries to ensure each of the nine experiments is at the same percentage SOC.} For all three cathodes, we withhold the fifth dataset as testing data, to verify that the KA-CRNN is not overfitting the training data. All training data, with testing data labeled in red, is available in Fig. {\ref{fig:data}}.

Returning to the physical system described in Eqs. \ref{eq:r1} through \ref{eq:r3}, we recall that certain reactions and parameters (phase change in the cathode material, R1 and R2) depend on the SOC, while others (electrolyte oxidation chemistry, R3) do not. To properly capture the true underlying chemical physics, the SOC-dependent parameters that define the two cathode phase transformations and the oxygen production rate (Eqs. \ref{eq:r1} and \ref{eq:r2}) {are defined uniquely in each cathode's $\bm{\theta}$} and optimized only to that cathode's DSC data. Meanwhile, the SOC-invariant parameters for the electrolyte oxidation are defined once and optimized across all cathode datasets (Eq. \ref{eq:r3}). Table \ref{tab:params} lists all learnable parameters, and whether they are cathode-specific KAN activations or universal scalar values. {To summarize, the set of all learnable KA-CRNN model parameters comprises two ``types'' of parameters: one $\bm{\theta}$ exists for each cathode, where encoded therein are individual arrays $\bm{\theta}_i$ ($i=1:11$) that, as per Eq. {\ref{eq:ChebyKAN}}, encode the response of each of the eleven R1 and R2 parameters in the top row of Table {\ref{tab:params}} to the SOC. The remaining parameters are the SOC-invariant and thus scalar-valued $n_3, A_3, E_{a,3}, b_3, \Delta H_3$ included in the second row of Table {\ref{tab:params}}. The overall pipeline used for evaluation and training is shown in Fig. {\ref{fig:fig1}}(E).}

% Requires: \usepackage{graphicx}
\begin{table}[h]
    \centering
    \renewcommand{\arraystretch}{1.5}
    \caption{Classification of all learnable parameters. }
    \begin{tabular}{cc}
        \hline
        \textbf{Inference technique} & \textbf{Parameters} \\ \hline
        Cathode-specific KAN activation. Function of SOC. & $n_1,  n_2, A_1, A_2, E_{a, 1}, Ea_{a, 2}, b_1, b_2, \Delta H_{1}, \Delta H_{2}, \nu$ \\ [-1ex]
        $[$cathode decomposition and O\textsubscript{2} production$]$ & \\
        Universal scalars. SOC-invariant, cathode-invariant. & $n_3, A_3, E_{a,3}, b_3, \Delta H_3$ \\ [-1ex]
        $[$electrolyte oxidation/combustion$]$&\\[-3ex] \\  \hline
    \end{tabular}
    \label{tab:params}
\end{table}

Optimization of all kinetic parameters to multiple datasets presents a challenge, especially when coupling cathode-specific and SOC-varying reactions to the universal electrolyte oxidation through $\nu$. To ensure a meaningful converged model, the loss function takes an augmented mean squared error (MSE) formulation with physics-informed penalties,

\begin{align}
    &\mathcal{L}(\bm{\theta})= \frac{1}{k}\sum_k \vert{Q_{tot,k}^{data}} - Q_{tot,k}^{KA-CRNN} \vert + \mathcal{L}_{mono}(\bm{\theta}) + \mathcal{L}_{min}(\bm{\theta}) + \mathcal{L}_{max}(\bm{\theta}), \label{loss_general} \\
    &\mathcal{L}_{mono}(\bm{\theta}) = \sum_{j=2}^9 \big[\text{Max}(\Delta H_2(\text{SOC}_{j-1}, \bm{\theta}) - \Delta H_2(\text{SOC}_{j}, \bm{\theta}), 0)   \label{loss_mono}\\   & \ \ \ \ \ \ \ \ \ \ \ \ \ \ \ \   +10^2 \ \text{Max}(\nu(\text{SOC}_{j-1}, \bm{\theta}) - \nu(\text{SOC}_{j}, \bm{\theta}), 0) \big],  \notag \\
    & \mathcal{L}_{min}(\bm{\theta}) =  \text{Max}\left(\frac{1}{n_3}-1, 0\right)  + 10\sum_{j=1}^9\sum_{k=1}^2  \text{Max}\left(\frac{1}{n_k(\text{SOC}_{j}, \bm{\theta})}-1, 0\right),   \label{loss_min} \\
    & \mathcal{L}_{max}(\bm{\theta}) =  0.05n_3^4 + b_3^4+  \sum_{j=1}^9  \sum_{k=1}^2 \big[ 0.05 n_k^2(\text{SOC}_{j}, \bm{\theta}) + b_k^3(\text{SOC}_{j}, \bm{\theta})  \big]. \label{loss_max}
\end{align}
$\mathcal{L}(\bm{\theta})$ is the overall loss function that informs model optimization through gradient stepping of the values in $\bm{\theta}$. In addition to the MSE term computed against the DSC data, $\mathcal{L}_{mono}(\bm{\theta})$ provides a soft monotonic penalty on the spinel to rock salt reaction enthalpy and the oxygen production stoichiometric coefficient, encouraging but not requiring both to increase across SOCs based on our physical knowledge from Sec. \ref{sec:method-review} of decreased thermal stability at elevated SOCs. These two variables are highly correlated via their overlapping behavior in the second observable heat release peak at all SOCs (Fig. \ref{fig:data}), making their inference slightly underconstrained and requiring this $\mathcal{L}_{mono}(\bm{\theta})$ penalty to avoid overfitting. {The monotonic penalty is achieved numerically in Eq. {\ref{loss_mono}} by computing the R2 reaction enthalpy and oxygen production stoichiometry at all training/testing SOC locations $j$, and penalizing the loss function at any point where the value at $j-1$ exceeds the value at $j$ (i.e., the parameter is not increasing). A $10^2$ weight is included on the $\nu$ penalty to bring it to the same scale as the naturally larger $\Delta H_2$ term.} A similar constraint for the layered to spinel reaction enthalpy is not needed, as those reaction enthalpies are fully constrained by the area under the first observable heat release peak in Fig. \ref{fig:data}, with no competing reactions. $\mathcal{L}_{min}(\bm{\theta})$ provides an increasing penalty on reaction orders less than unity as they approach zero, which helps maintain reasonable rate law formulations and ensure numerical stability. {The shape of this penalty, as shown in Eq. {\ref{loss_min}}, is defined by the reciprocal function. The subtraction and maximum functions serve to smoothly reduce the penalty to zero as the reaction orders approach and exceed unity. The nested sum applies this penalty at all nine train/test SOC $j$, as well as to both KA-CRNN reaction orders for reactions $k=1$ and $k=2$. $n_3$ is treated separately, as is it a scalar value that does not require SOC sampling.} $\mathcal{L}_{max}(\bm{\theta})$ similarly provides an increasing penalty for large values of all reaction orders and non-exponential temperature factors, to prevent nonsensical parameterizations. This is especially relevant for the electrolyte oxidation reaction, which given the very narrow peaks at high SOCs is prone to drifting toward extreme parameter values. {The scales of these penalties are adjusted with weights in front of each term in Eq. {\ref{loss_max}}, while the rate of increase is adjusted with integer exponents. As in Eq. {\ref{loss_min}}, $n_3$ and $b_3$ are treated independently, while $n_k$ and $b_k$ for $k=1:2$ are treated in the nested sampling loops to account for the two reactions at nine SOCs.} These loss constraints provide knowledge-based guardrails that ensure a meaningful final model without hampering its interpretability or realistic chemical physics.

KA-CRNN predictions are computed by integrating the governing equations, as parameterized by the KA-CRNN, in the DifferentialEquations package in Julia \cite{rackauckas_differentialequationsjl_2017}.  Model optimization is carried out by updating model parameters according to the gradients of the loss function as computed via autodifferentiation in the ForwardDiff package \cite{revels_forward-mode_2016} in Julia. {Optimization of the R3 parameters is handled identically as in prior CRNN work, where the network parameters directly encode the Arrhenius and mass action parameters and thus network parameter optimization directly tunes the real parameter values. For the $\bm{\theta}$-encoded KA-CRNN parameters of R1 and R2, meanwhile, network gradient updates are performed on the individual $w^{\psi}_{i,n}$ values comprising each cathode's $\bm{\theta}$, which implicitly updates the SOC-varying real parameters as per Eq. {\ref{eq:ChebyKAN}}.} The ADAM optimizer is used for training \cite{kingma_adam_2017}, with a learning rate of $5\times10^{-4}$.

\section{Results and Discussion}
\subsection{SOC-varying DSC predictions with sparse KA-CRNN model}
\begin{figure}
\centering
\includegraphics[width=1.0\linewidth]{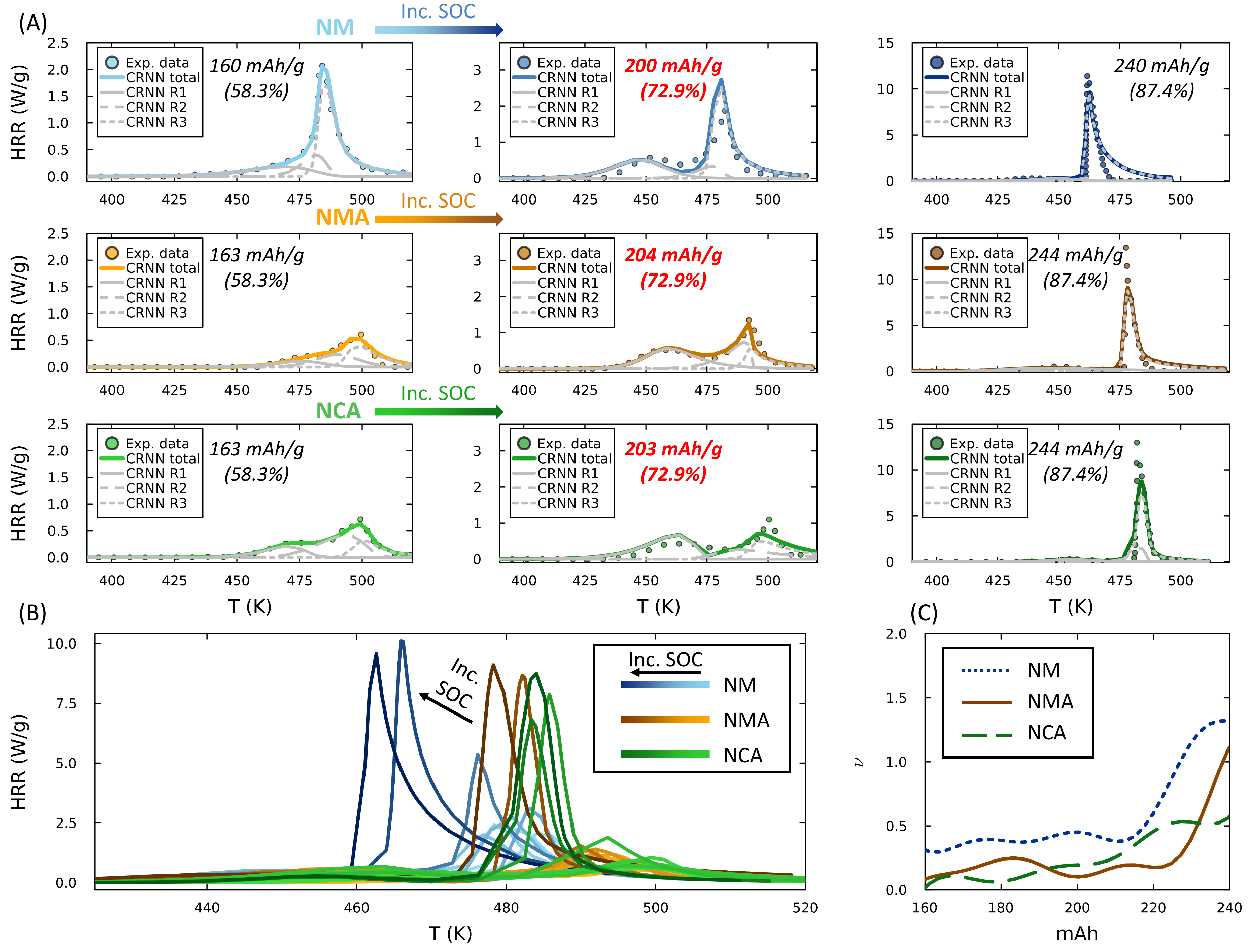}
	\caption{(A) DSC reconstructions for all three studied cathodes (NM, NMA, and NCA, from top to bottom), at three example SOCs (minimum, testing, and maximum from left to right). SOCs are labeled in mAh/g (black for training, red for testing). Note y-axes are scaled to each column for visibility.
    (B) All nine SOCs for all three cathodes plotted together, showing significantly distinct trends in peak heat release, peak heat release temperature, and total heat release as the SOC increases. (C) Oxygen evolution stoichiometric coefficient KA-CRNN activations for each cathode material, highlighting critical SOCs around 210 mAh/g to 230 mAh/g across all materials.}
	\label{fig:Recon}
\end{figure}

Reconstructed DSC profiles for the NM cathode plus electrolyte experiments at varying SOCs \cite{cui_navigating_2025} are shown in the top blue row of Fig. \ref{fig:Recon}(A). Inspection of the three provided training profiles with appropriately scaled y-axes indicates strong performance, with the KA-CRNN's learned responses to the SOC for the two cathode decomposition reactions (R1 and R2) enabling them to change in location, magnitude, and shape as the SOC increases. Also observable are significant changes in the shape, size, and peak temperature of the R3 peak at increasing SOCs, which are notable given its fixed kinetic parameters across all SOCs (and across all cathodes). Instead, these shifts demonstrate the successful implementation of two key behaviors in the KA-CRNN: first, the SOC-sensitive oxygen production stoichiometric coefficient (shown in Fig. \ref{fig:Recon}(C)) increases the total oxygen availability at higher SOCs as is well-reported in the literature, enabling R3 to release more heat even with a fixed per-unit-oxygen reaction enthalpy $\Delta H_3$. Secondly, the limiting oxygen reactant for R3 is produced directly during R2, making R2 initiation a prerequisite for the onset of R3 heat release, and thus making the shape of the R3 heat release peak dependent on the shape of the R2 phase change peak. These two pathways give the KA-CRNN the expressivity needed to control, respectively, the size and onset of R3 through the SOC-dependent R2 kinetics. 

Inspection of the blue NM curves in the combined Fig. \ref{fig:Recon}(B) further emphasizes the stark difference in peak heat release, total heat release, and onset temperature as the SOC changes. The learned KA-CRNN activations are successful at capturing this shifting behavior across all SOCs well. The oxygen production stoichiometric coefficient's variation with respect to the SOC is shown in Fig. \ref{fig:Recon}(C), where between 210 mAh/g and 220 mAh/g there is a sudden jump in the blue NM profile corresponding to the sudden increase ($>2\times$) in peak heat release in the reconstructed profiles at those SOCs. This visualization in Fig. \ref{fig:Recon}(C) is available with zero postprocessing thanks to the direct KA-CRNN activation function learning via polynomial basis functions, facilitating highly interpretable results that not only fit the data but also correspond well to the realistic underlying physics. The remaining KA-CRNN parameters (see Table \ref{tab:params} for the complete list) can be visualized similarly, and are discussed in the next section. Finally, we remark that the ability of the KA-CRNN to fit the testing data in the top row of Fig. \ref{fig:Recon}(A) well (200mAh/g, labeled in red) despite no nearby training points is a significant result for this approach, and emphasizes the continuous yet overfitting-resistant nature of its response to the SOC. The fixed electrolyte oxidation parameters learned across this and the remaining NMA and NCA data are reported in Table \ref{tab:elec_oxy_param}.

Next, the same results are shown for the NMA cathode plus electrolyte DSC data in the orange middle row of  Fig. \ref{fig:Recon}. Similar conclusions can be drawn: the responses of R1 and R2 to the SOC are both learned well, and adjust in shape and magnitude as needed to properly fit all training data. R3, despite its SOC-invariant chemistry that is shared with the NM and NCA cases (see parameters in Table \ref{tab:elec_oxy_param}), is still shown to capture all of the experimental peaks well thanks to its dependence on the SOC-sensitive R2 for the reactant oxygen production. The relatively new NMA chemistry \cite{li2020high} is more stable than the cobalt-free NM cathode, which manifests in a lower stoichiometric coefficient at all SOCs as well as a higher critical SOC in the orange curve of Fig. \ref{fig:Recon}(C), leading to less overall electrolyte oxidation and R3 heat release.

The KA-CRNN's ability to reconcile these significant differences between the NMA and NM cathode stabilities even with the constant scalar R3 parameters is notable, and provides interesting kinetic insights. R2 in the NM reconstructions (top blue row), especially at low SOCs, is relatively narrow. All oxygen is produced relatively early by the narrow R2, allowing the Arrhenius rate law for R3 to largely control the electrolyte oxidation behavior. In the NMA cathode, however (middle orange row), the R3 peak occurs at a higher temperature in the data, thanks to the more stable chemistry \cite{li2020high}. The scalar R3 parameters at these higher temperatures predict an exponentially faster reaction due to the Arrhenius rate law, which is not corroborated by the still-broad DSC peaks (especially at low SOCs). To correct for this, the KA-CRNN optimization learns a broader R2 peak for the NMA cathode, evolving smaller quantities of oxygen over a wider temperature window. This wider SOC-dependent R2 peak effectively broadens the high Arrhenius rate R3 peak through a mass action law limitation that restricts oxygen availability. Such discussion highlights a unique capability of the KA-CRNN framework: the KAN activations enable continuous SOC-based modeling, while the strictly physics-enforced CRNN backbone ({via the Arrhenius and mass action parameterization}) facilitates such direct kinetic interpretation of the real model results. {Further plotting of the actual oxygen production and consumption rates is available in Appendix A to further illustrate this point.}

Extrapolation to the testing data for the NMA cathode is also seen to be strong in Fig. \ref{fig:Recon}(A), as is the ability of the oxygen evolution stoichiometric parameter to capture the critical SOC and accompanying sudden heat release in Fig. \ref{fig:Recon}(C). 

We finally demonstrate reconstructed DSC results for the NCA cathode plus electrolyte DSC data in the bottom green row of Fig. \ref{fig:Recon}, offering comparison with a more traditional cathode chemistry. Similarly strong performance is seen in the training reconstructions. The testing reconstruction is not quite as accurate as in the first two cathode materials, but it retains good accuracy and peak location despite the sparse training data. Inspection of all overlaid reconstructions in Fig. \ref{fig:Recon}(B) enables comparison of overall thermal stability trends, where we note that the NCA and NMA are quite similar in their improved thermal stability over the NM. Both release significantly less heat than the NM cathode at lower SOCs, and generally have higher peak heat release temperatures, indicating higher thermal stability. The NCA suffers from a slightly lower critical SOC than the NMA, as can be seen in Fig. \ref{fig:Recon}(C), while the NMA's highest SOC samples suffer from slightly lower peak heat release temperatures (Fig. \ref{fig:Recon}(B)). Full sets of training data reconstructions (i.e., Fig. \ref{fig:Recon}(A) at all SOCs for all cathodes) are available in Appendix C.

\begin{table}
    \centering
    \caption{Learned electrolyte oxidation parameters}
    \begin{tabular}{ccccc}
         ln$A_3$ &  $E_{a,3}$ (J/mol)&  $b_3$&  $\Delta H_3$ (J/g)& $n_3$\\ \hline \\ [-2.5ex]
         $132.45$&  5.37 $\times$ 10$^5$ &  -0.069 &  1779.39 & 3.250\\
    \end{tabular}
    \label{tab:elec_oxy_param}
\end{table}

\subsection{KA-CRNN parameter visualizations and SOC trends}

\begin{figure}
\centering
\includegraphics[width=0.95\linewidth]{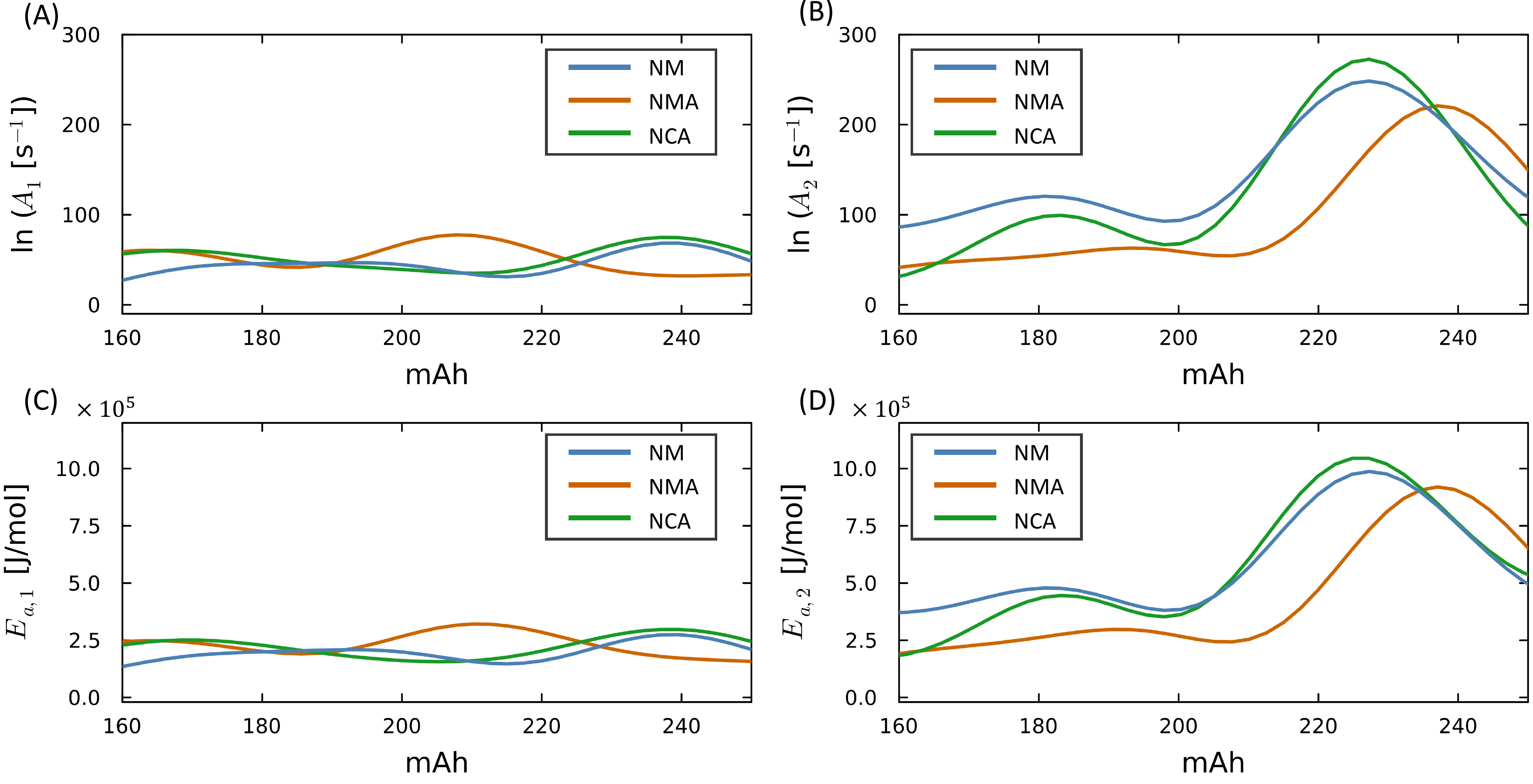}
	\caption{Learned KA-CRNN activations for the two frequency factors (A-B) and two activation energies (C-D) of all three cathode materials. Note increase in ln$A_2$ and $E_{a, 2}$ occurs at a higher SOC for NMA cathode, corresponding to its higher critical SOC. KA-CRNN activations and discussion for all remaining kinetic parameters are available in Appendix B. }
	\label{fig:Acts_all}
\end{figure}
The learned DSC reconstructions of Fig. \ref{fig:Recon} resulted in three separate sets of SOC-dependent KA-CRNN kinetic parameters for R1 and R2, as well as a single set of scalar-valued parameters for R3. The latter are shown in Table \ref{tab:elec_oxy_param}, while we plot the KA-CRNN Arrhenius parameters for all three cathodes in Fig. \ref{fig:Acts_all} (remaining activations are available in Appendix B). The first phase change reaction's parameters (Figs. \ref{fig:Acts_all}(A) and (C)) see modest changes across the SOC, which when compared to the fairly consistent R1 behavior seen in the data of Figs. \ref{fig:data} and \ref{fig:Recon} is to be expected. The second phase change reaction's parameters (Figs. \ref{fig:Acts_all}(B) and (D)) see significant shifts just before each cathode material's critical SOC, once again corroborating the sudden increase in heat release seen in the underlying DSC data of Figs. \ref{fig:data} and \ref{fig:Recon}, the oxygen stoichiometric parameter's behavior in Fig. \ref{fig:Recon}(C), and the dependence of R3 heat release on R2 activity. These results demonstrate that the KA-CRNN framework is able to directly learn the underlying chemical behaviors that explain these responses in a visualizable and fully self-consistent manner, in addition to accurately describing the continuous thermal responses of these components to the SOC. {We additionally note relatively large values for the pre-exponential factor and activation energy of the electrolyte oxidation reaction in Table {\ref{tab:elec_oxy_param}}. Consulting the highly stiff behavior exhibited by the experimental data at the maximum state of charge in Fig. {\ref{fig:Recon}(A)}, we find these large values in accordance with expectations and with experiments. Future work to better resolve the electrolyte oxidation reaction with a potentially multi-step pathway could help to further define and constrain these values.} Further discussion and visualization of the complete set of activations for all kinetic parameters across all cathodes is available in Appendix B. The numerical values for the continuous parameters are available in the supplementary material.

We finally emphasize that, thanks to the CRNN coupling, the KA-CRNN activation functions learned here are sampled at any SOC into real kinetic parameters used in the Arrhenius and mass action formulation of Eq. \ref{eqn_rate_log}. Downstream application of the models developed in this work requires no black-box networks, and needs only to refer to the simple activation functions of Figs. \ref{fig:Acts_NM} through \ref{fig:Acts_NCA} (available in the supplementary materials), which when sampled at any SOC can be integrated forward in time identically to any other thermal-kinetic model.

\subsection{Future outlook}

The current study provides a first-of-its-kind model describing the continuous SOC variation of the cathode and electrolyte submodel during thermal runaway with a fully {interpretable network structure parameterized by the real Arrhenius and mass action laws}. {Future studies on similar datasets could involve further battery components such as the anode, SEI, and separator; as well as further side reactions and effects such as the anode-electrolyte exothermic interactions, lithium plating, and SEI decomposition. Such studies would facilitate complete, cell-level thermal runaway kinetic models with physics-based, continuous SOC variation that enable live safety evaluation while retaining strict adherence to fundamentally understood reaction pathways and component interaction effects. A blueprint for such a scale-up is available in recent work {\cite{koenig_comprehensive_2025}} that investigated uncertain modeling of the pipeline from component-scale experimental data to full-cell thermal runaway models leveraging the Bayesian CRNN approach {\cite{li_bayesian_2023}}, which we note is fully integrable with KA-CRNNs thanks to the overarching CRNN ecosystem. Bayesian KA-CRNNs could further advance the field and enable fully descriptive, probabilistic models that account for experimental uncertainty as well as SOC variation at the cell scale.}

Thanks to the prior knowledge encoded directly into the reaction model, the currently proposed framework was able to extract a significant amount of realistic model behavior from limited heat release DSC data only, including oxygen evolution stoichiometry. That said, we recognize as well that augmented datasets could aid in further model fidelity. Combined DSC and TGA datasets, for example, could serve to better pin down the oxygen evolution kinetic behavior through mass change loss terms, while the incorporation of mass spectrometry data could also lead to models that continuously report evolved gas species and ratios across SOCs. {Further, we remark that while the global R3 reaction was able to largely exploit shared information across all three datasets to place the exothermic R3 peaks in highly variable yet highly accurate temperature locations in Fig. {\ref{fig:Recon}(A)}, we acknowledge that at the highest SOCs the model was not able to fully reconstruct the full heat release peak values. While a trivial numerical fix might be to split R3 into three separate reactions (one per cathode), this would violate key aspects of the literature-informed reaction pathway development provided in Sec. {\ref{sec:methods}}. Future effort here could potentially aim to further resolve the global electrolyte oxidation physics through additional reactions, pathways, or parameterizations while retaining an honest and accurate interpretation of previous fundamental work in the literature.} Nonetheless, the current results already demonstrate the capability of KA-CRNNs to learn SOC-varying thermal runaway kinetic models from fairly sparse data thanks to their incorporation of the cathode-electrolyte reaction physics across a continuous range of SOCs, retaining full interpretability of all reactions and tracked species while adhering fully to previously reported chemical pathways.

\section{Conclusions}
This work employed Kolmogorov–Arnold Chemical Reaction Neural Networks to construct state-of-charge dependent kinetic models for cathode decomposition and electrolyte oxidation in lithium-ion batteries, delivering unified predictive models for NM, NMA, and NCA chemistries that remain strongly grounded in realistic chemical physics. A detailed review of the impacts of the SOC on cathode oxygen evolution and electrolyte oxidation provided realistic reaction pathways based on prior experimental results, while a carefully tailored hybrid KA-CRNN structure encoded these pathways into a learnable, interpretable network. KA-CRNN models trained on recently reported experimental data were able to accurately capture all relevant chemical pathways including critical SOCs, oxygen production and electrolyte oxidation, heat release rates, and peak temperature values, with strong generalization capabilities highlighted by validation on withheld testing data at an unseen SOC. The delivered models are able to predict the continuous dependence of these exothermic cathode and electrolyte interactions on a cell's SOC, potentially facilitating finer and higher-resolution safety monitoring in practical applications, especially under SOC-limited safety regimes where previous 100\% SOC models are not applicable.

A key novelty of this approach is the continuous response of the KA-CRNN kinetic parameters to the SOC, a significant departure from standard approaches that use fixed parameters at single or discrete SOCs only. Additionally, the modularity of KA-CRNNs with generic CRNNs and standard rate forms enables the simultaneous inference of a shared, SOC-invariant electrolyte oxidation reaction, providing a more generalizable model with sparser and meaningful parameters that facilitates simpler and more robust training through shared information across different chemistries. KA-CRNNs are a general approach for the modeling and elucidation of unknown external factors through physically-grounded kinetic models, and are not limited to modeling the effects of SOC. In the battery thermal runaway field especially, where there are many global and single-step reactions with murky dependencies on external variables that could include pressure and electrochemical operating conditions, KA-CRNNs show promise in elucidating such relationships and providing accurate, continuous, and interpretable predictive models.

\section{Declaration of Competing Interest}
The authors declare that they have no known competing financial interests or personal relationships that could have appeared to influence the work reported in this paper.

\section{Acknowledgements}
SD acknowledges the support from the National Science Foundation under Grant No. CBET-2143625. BCK is partially supported by the National Science Foundation Graduate Research Fellowship under Grant No. 1745302. This research was supported by the Department of Energy (DOE) under Grant No. DE-EE0011364. The authors would like to thank Professor Martin Bazant for his expertise on cathode phase transformation theory.

\section*{CRediT authorship contribution statement}
\textbf{Benjamin C. Koenig:} Conceptualization, Methodology, Investigation, Software, Formal Analysis, Writing - Original Draft. \textbf{Sili Deng:} Conceptualization, Project Administration, Writing - Review and Editing. 

\bibliographystyle{elsarticle-num} 
\bibliography{KA_CRNN.bib}

\clearpage
\renewcommand\thetable{A\arabic{table}}
\renewcommand\thefigure{A\arabic{figure}}
\renewcommand\theequation{A\arabic{equation}}
\setcounter{equation}{0} 

\setcounter{figure}{0} 
\setcounter{table}{0}
\setcounter{section}{0} 

\section*{Appendix A. Oxygen production and consumption rates}

\begin{figure}
\centering
\includegraphics[width=0.95\linewidth]{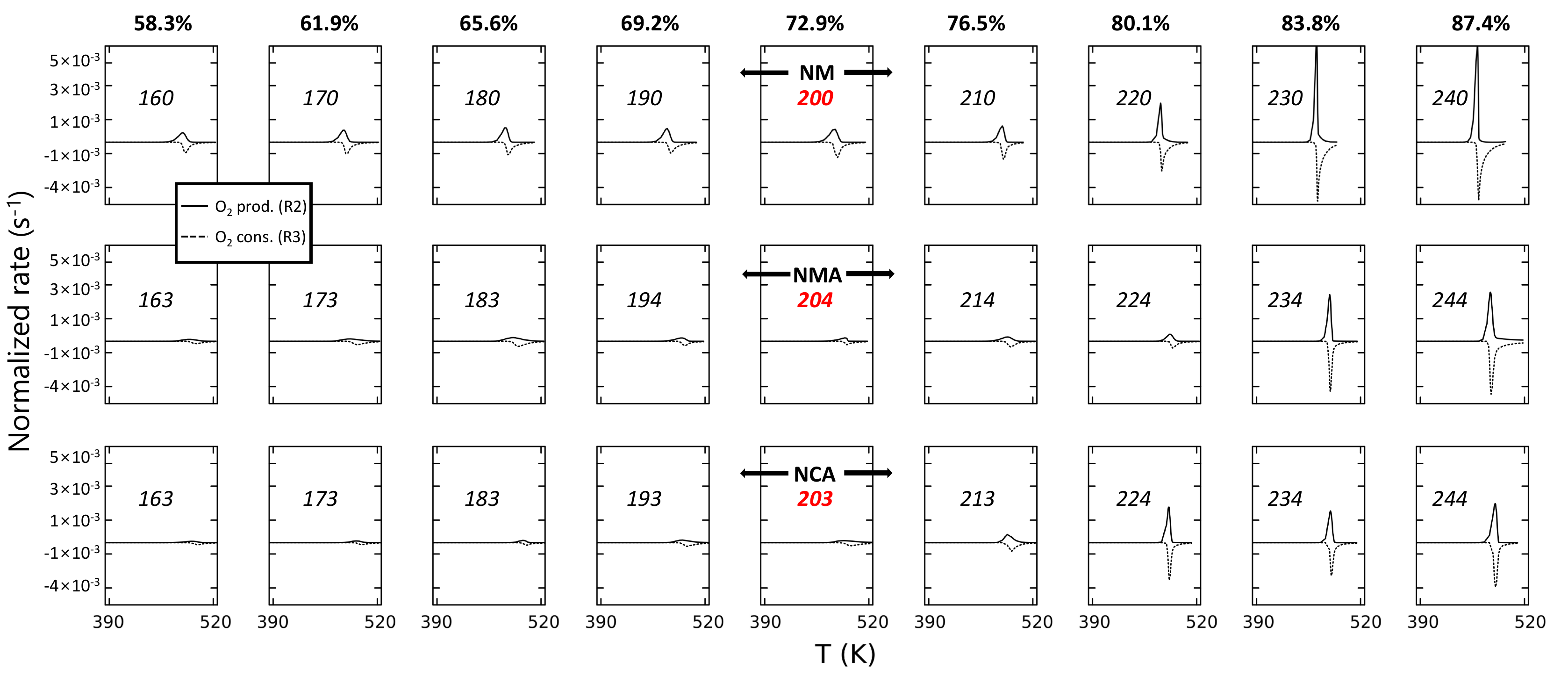}
	\caption{{Oxygen production and consumption rates, as computed in Eq. {\ref{eqn_species_rate_3}}. Trends match up with the governing stoichiometric coefficient $\nu$ shown in Fig. {\ref{fig:Recon}}(C). Relative changes in shape, both horizontally across SOCs and vertically across cathodes, highlight the SOC-varying control that R2 exerts on R3. Italicized numbers in plots are mAh/g SOCs, while percent values above each column are the fixed lithium removal levels.}}
	\label{fig:appendix_oxy}
\end{figure}

{Here we briefly present Fig. {\ref{fig:appendix_oxy}}, where oxygen production and consumption rates are plotted across all datasets and all cathodes. The expected increase with SOC is observed, as well as generally poorer thermal stability in the NM cathode as was initially observed in Fig. {\ref{fig:Recon}}. The fully resolved production and consumption plots here help to contextualize the way in which the SOC-controlled R2 production rates (solid lines, positive) effectively encode SOC-varying behavior in the R3 consumption rates (dashed lines, negative) despite the R3 parameters themselves being SOC-invariant.}

\renewcommand\thetable{B\arabic{table}}
\renewcommand\thefigure{B\arabic{figure}}
\renewcommand\theequation{B\arabic{equation}}
\setcounter{equation}{0} 

\setcounter{figure}{0} 
\setcounter{table}{0}
\setcounter{section}{0}

\section*{Appendix B. KA-CRNN activations for NCM and NM cathodes}
\begin{figure}
\centering
\includegraphics[width=0.8\linewidth]{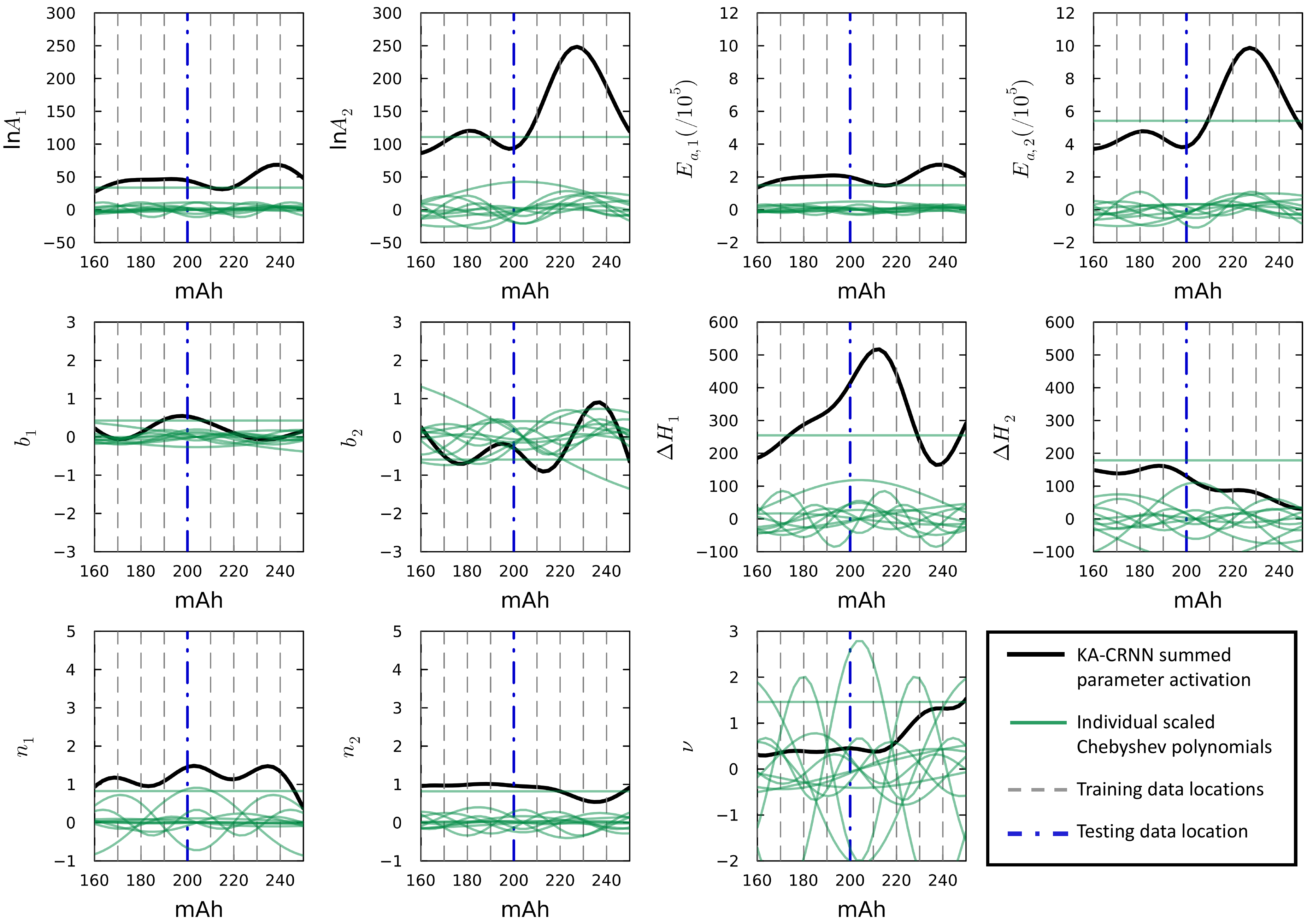}
	\caption{Learned KA-CRNN activations for the NM cathode. Individual green Chebyshev polynomials sum to the black KA-CRNN parameter functions. }
	\label{fig:Acts_NM}
\end{figure}

\begin{figure}
\centering
\includegraphics[width=0.8\linewidth]{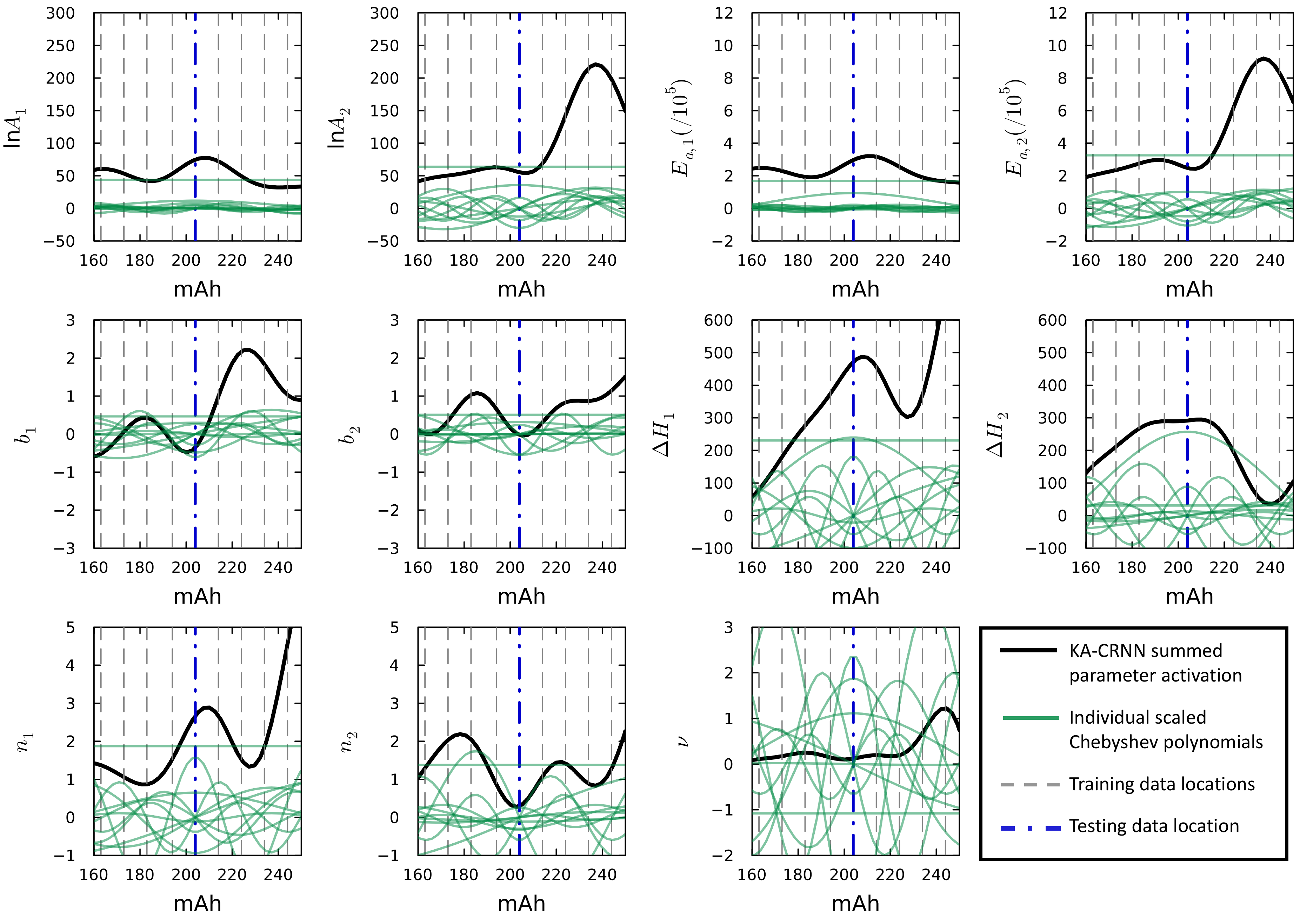}
	\caption{Learned KA-CRNN activations for the NMA cathode. Individual green Chebyshev polynomials sum to the black KA-CRNN parameter functions. }
	\label{fig:Acts_NMA}
\end{figure}

\begin{figure}
\centering
\includegraphics[width=0.8\linewidth]{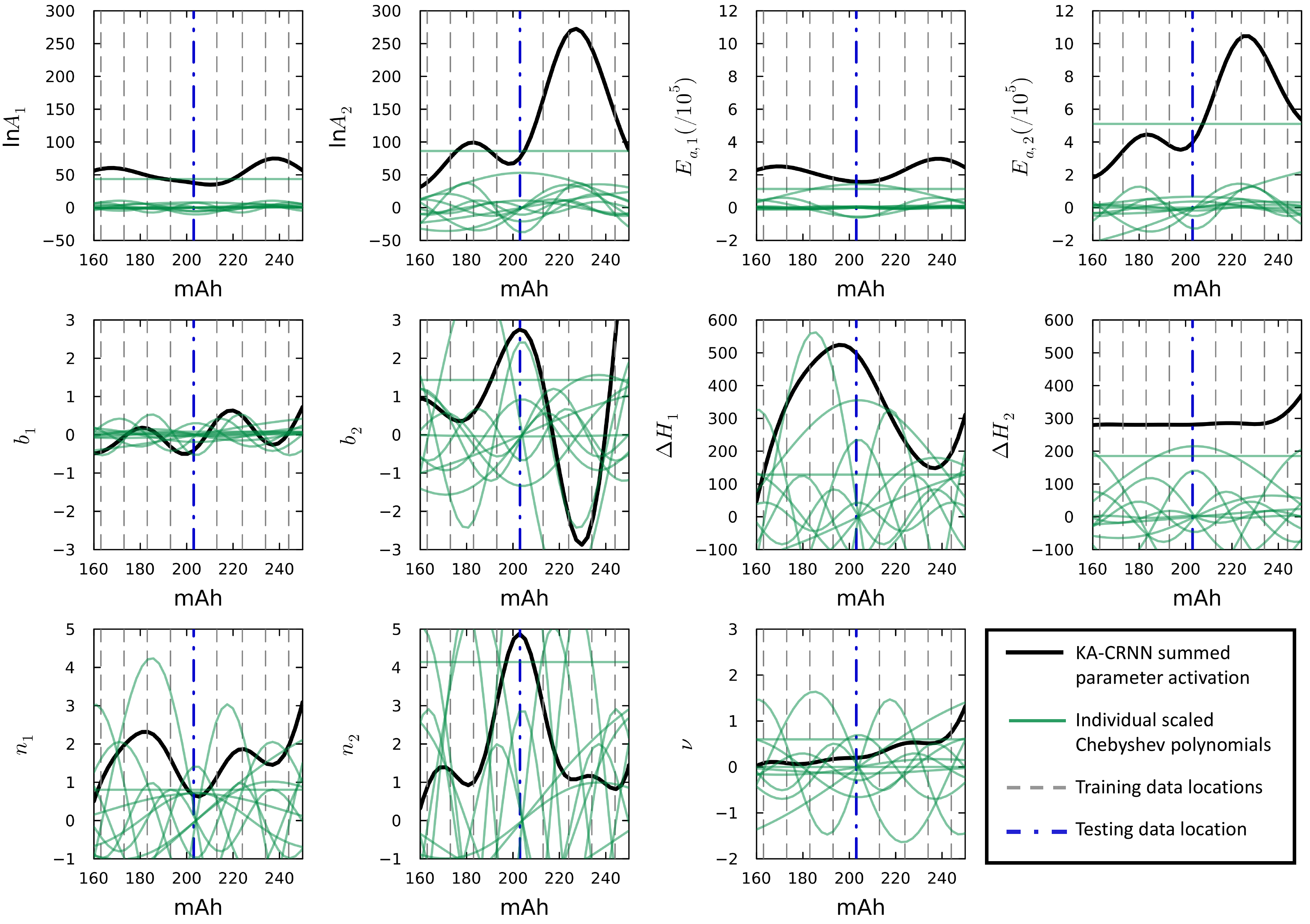}
	\caption{Learned KA-CRNN activations for the NCA cathode. Individual green Chebyshev polynomials sum to the black KA-CRNN parameter functions. }
	\label{fig:Acts_NCA}
\end{figure}

KA-CRNN activations are provided here for all kinetic parameters of all three studied cathodes (Figs. \ref{fig:Acts_NM}, \ref{fig:Acts_NMA}, and  \ref{fig:Acts_NCA}). Comparison across these activations reveals similar qualitative trends despite the differences across cathodes in the reconstructed DSC results in Fig. \ref{fig:Recon}. A key behavior not previously discussed in Fig. \ref{fig:Acts_all} is that of $\Delta H_1$. In all three cathodes, it follows a trend of increasing enthalpy up to 200 mAh/g, then a noticeable decrease toward the critical SOC. The decrease toward higher SOCs appears counterintuitive, as it is well known (and also demonstrated in the current work) that the thermal stability of these cathode materials decreases with increased SOCs. Interestingly, a close inspection of the training data reveals that R1 does not appear to follow this otherwise well-verified trend, and in fact in many cases releases more total heat at intermediate SOCs, rather than at the critical SOC and above. For the NM cathode, for example, the R1 peak heat release rate at 210 mAh/g is 33\% higher than that at 220 mAh/g, while for the NCA cathode the 213mAh/g peak is 21\% higher than the 224 mAh/g peak. It is not clear from the current results whether this is a true chemical behavior or an artifact of the experimental methods, but in either case it is clear that the behavior learned by the KA-CRNN, whether expected or unexpected, can be traced directly to behaviors present in the underlying training data.

\renewcommand\thetable{C\arabic{table}}
\renewcommand\thefigure{C\arabic{figure}}
\renewcommand\theequation{C\arabic{equation}}
\setcounter{equation}{0} 

\setcounter{figure}{0} 
\setcounter{table}{0}
\setcounter{section}{0} 

\section*{Appendix C. KA-CRNN data reconstructions at remaining states of charge}

\begin{figure}
\centering
\includegraphics[width=0.87\linewidth]{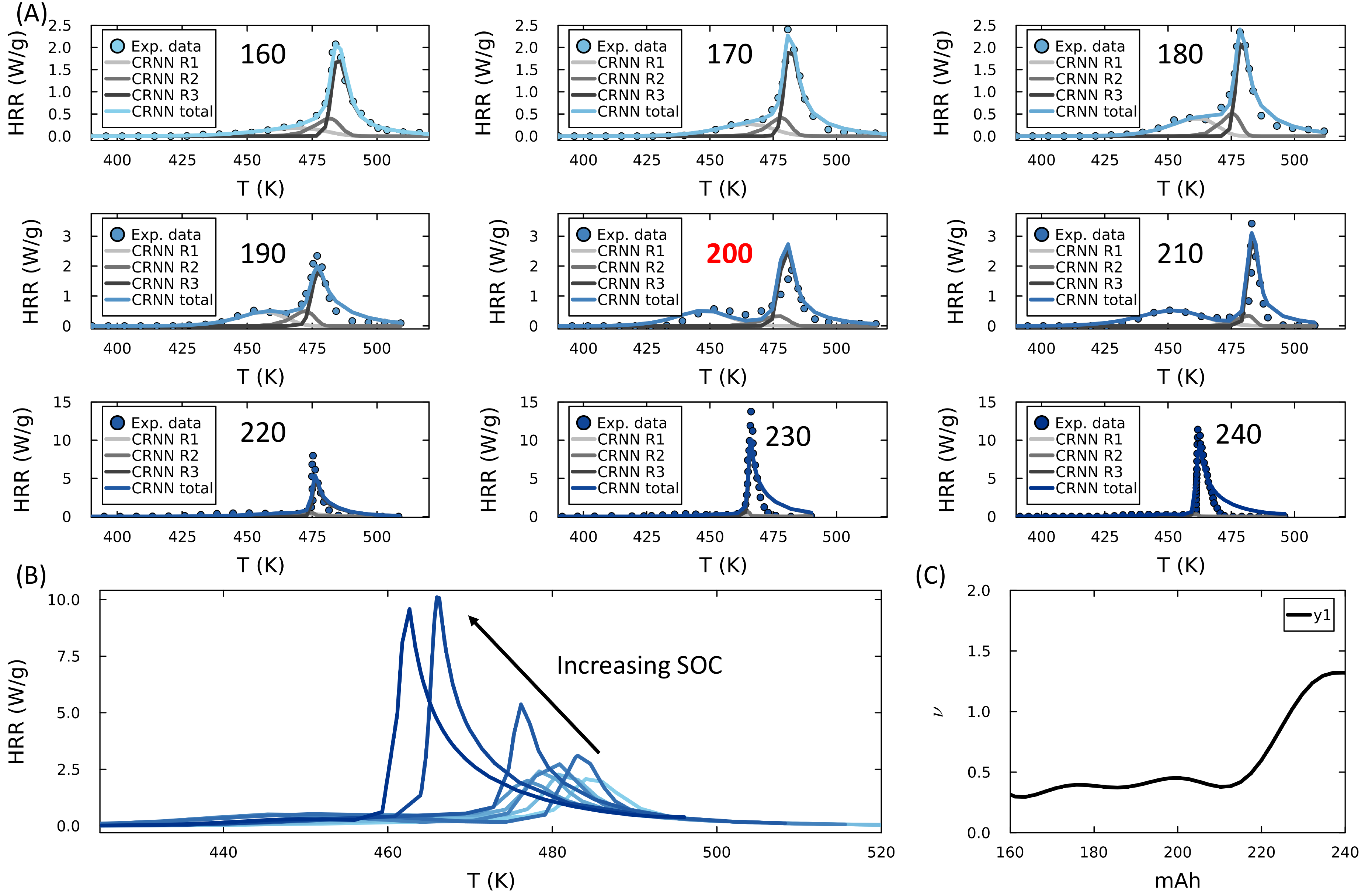}
	\caption{(A) NM DSC reconstructions at all SOCs (train/test). (B) All reconstructions plotted together to visualize SOC effect. (C) Oxygen stoichiometric coefficient KA-CRNN activation, highlighting critical SOC.}
	\label{fig:Recon_NM}
\end{figure}

\begin{figure}
\centering
\includegraphics[width=0.87\linewidth]{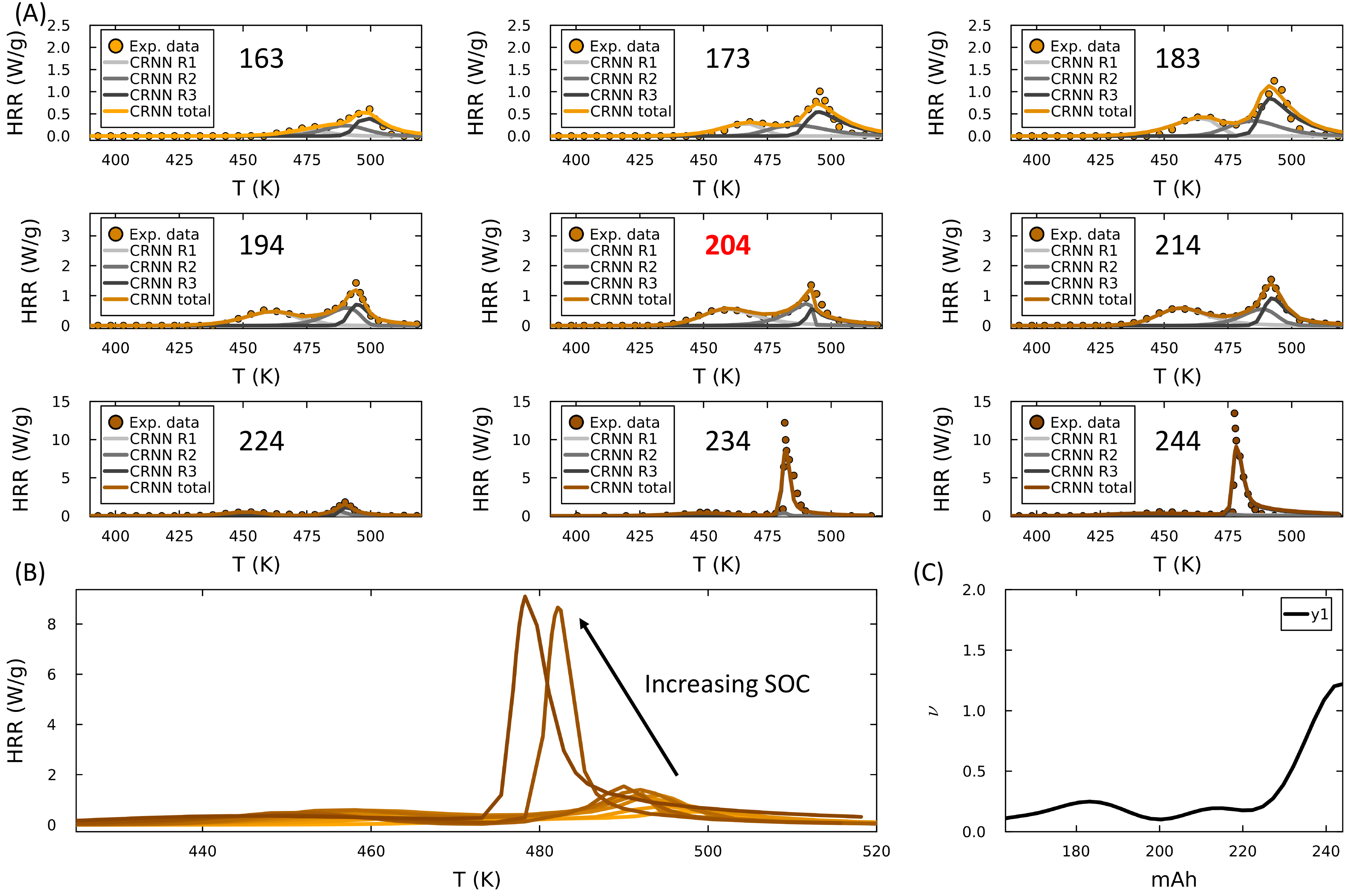}
	\caption{(A) NMA DSC reconstructions at all SOCs (train/test). (B) All reconstructions plotted together to visualize SOC effect. (C) Oxygen stoichiometric coefficient KA-CRNN activation, highlighting critical SOC.}
	\label{fig:Recon_NMA}
\end{figure}

\begin{figure}
\centering
\includegraphics[width=.87\linewidth]{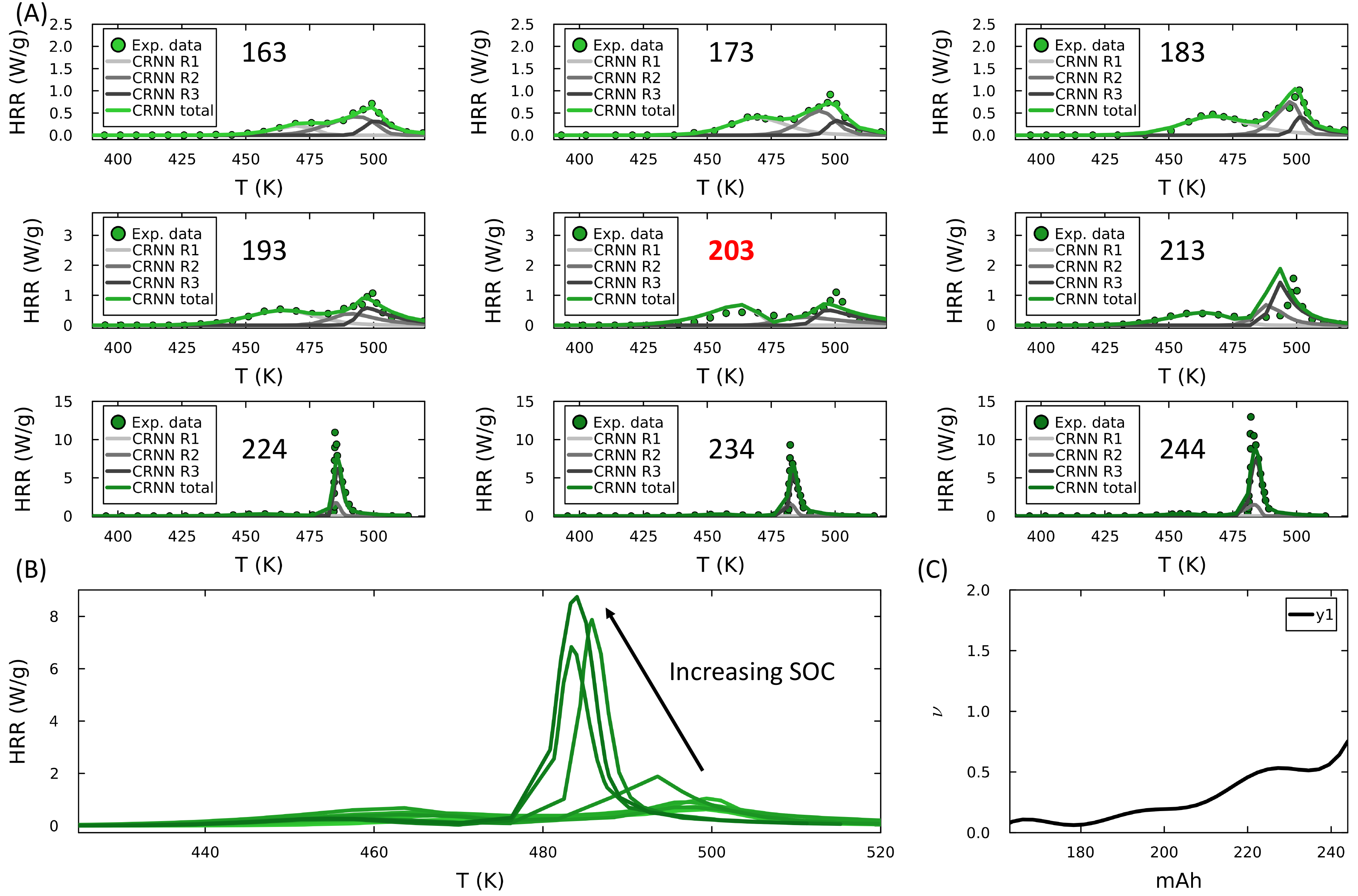}
	\caption{(A) NCA DSC reconstructions at all SOCs (train/test). (B) All reconstructions plotted together to visualize SOC effect. (C) Oxygen stoichiometric coefficient KA-CRNN activation, highlighting critical SOC.}
	\label{fig:Recon_NCA}
\end{figure}

The results in Fig. \ref{fig:Recon} show only three reconstructed data profiles for each cathode, to facilitate concise discussion. In this appendix we report all reconstructed data profiles: the NM cathode in Fig. \ref{fig:Recon_NM}, NMA in Fig. \ref{fig:Recon_NMA}, and finally NCA in Fig. \ref{fig:Recon_NCA}.

\section*{Appendix D. Supplementary data}
Supplementary material is available in the published version of this article, available at \href{https://doi.org/10.1016/j.est.2026.121853}{10.1016/j.est.2026.121853}.

\end{document}